\documentclass[sigconf]{acmart} 
\pdfoutput =1
\usepackage{booktabs} 
\usepackage{xcolor}
\usepackage[ruled,vlined,linesnumbered,noresetcount]{algorithm2e}
\usepackage{enumitem}

\copyrightyear{2019}
\acmYear{2019} 
\setcopyright{iw3c2w3}
\acmConference[WWW '19]{Proceedings of the 2019 World Wide Web Conference}{May 13--17, 2019}{San Francisco, CA, USA}
\acmBooktitle{Proceedings of the 2019 World Wide Web Conference (WWW '19), May 13--17, 2019, San Francisco, CA, USA}
\acmPrice{}
\acmDOI{10.1145/3308558.3313634}
\acmISBN{978-1-4503-6674-8/19/05}

\begin{document}
\title[Firsthand Opiates Abuse on Social Media]{Firsthand Opiates Abuse on Social Media: Monitoring Geospatial Patterns of Interest Through a Digital Cohort}

\author{Duilio Balsamo}
\affiliation{%
  \institution{University of Turin \& ISI Foundation}
  \city{Turin}
  \state{Italy}
}
\email{duilio.balsamo@unito.it}

\author{Paolo Bajardi}
\affiliation{%
  \institution{ISI Foundation}
  \streetaddress{Via Chisola 5}
  \city{Turin}
  \state{Italy}
  \postcode{10126}
}
\email{paolo.bajardi@isi.it}

\author{Andr\'e Panisson}
\affiliation{%
  \institution{ISI Foundation}
  \streetaddress{Via Chisola 5}
  \city{Turin}
  \state{Italy}
  \postcode{10126}
}
\email{andre.panisson@isi.it}

\begin{abstract}
  
In the last decade drug overdose deaths reached staggering proportions in the US.  
Besides the raw yearly deaths count that is worrisome \textit{per se}, 
an alarming picture comes from the steep acceleration of such rate that increased by 21\% from 2015 to 2016. While traditional public health surveillance suffers from its own biases and limitations, digital epidemiology offers a new lens to extract signals from Web and Social Media that might be complementary to official statistics.
In this paper we present a computational approach to identify a 
digital cohort that might provide an updated and complementary view on the opioid crisis.
We introduce an information retrieval algorithm suitable to identify relevant subspaces of discussion on social media, 
for mining data from 
users showing explicit  interest in discussions about opioid consumption in Reddit. Moreover, despite the pseudonymous nature of the user base, almost 1.5 million users were geolocated at the US state level, resembling the census population distribution with a good agreement. A 
measure of prevalence of 
interest in opiate consumption has been estimated at the state level, 
producing a novel indicator 
with 
information that is not entirely encoded in the standard surveillance. Finally, we further provide a 
domain specific vocabulary containing informal lexicon and street nomenclature extracted by user-generated content
that can be used by researchers and practitioners to implement novel digital public health surveillance methodologies for supporting policy makers in fighting the opioid epidemic.

\end{abstract}

%
%



\maketitle

\section{Introduction}
The opioid crisis emerged in the US from the interplay of several determinants and evolved over time through three major phases: from the abuse of drug prescriptions of late 80's \cite{leung20171980}, to heroin injecting users soaring in 2010 \cite{michel2015shifting} and the very recent rise of synthetic opioids flooding the drug market~\cite{kolodny2015prescription}.
The rate of overdose deaths involving opioids shows heterogeneous geographical patterns ranging from 4.9 per 100,000 inhabitants in Texas to 43.4 in West Virginia in 2016 \cite{seth2018overdose}. Unfortunately, despite the effort of national agencies in monitoring the phenomenon, state-level estimates suffer biases due to unequal coverage of the surveillance system as well as intrinsic biases in counting overdose deaths that lead to low specificity of drugs involved \cite{Hedegaard,ruhm2017geographic,landen2003methodological}. Besides drug overdose maps, opioid prescription maps are provided on yearly basis by the Centers for Disease Control and Prevention (CDC), showing heterogeneous patterns also due to different prescribing policies implemented by individual states. In this perspective, the gold standard represented by official surveillance statistics has to be carefully considered in the light of the known biases of the reported numbers and estimates.
A complementary way to investigate these phenomena is by field studies, through ethnographic approaches and structured interviews: sadly these methods require an exceptional effort in order to gather insights from firsthand users \cite{mars2014every} or subjects under opioid substitution maintenance treatment \cite{
sordo2017mortality}, and are often limited to small sample sizes ranging from tens to hundreds of individuals. 
In this context, a digital epidemiology approach \cite{salathe2012digital} aimed at integrating and complementing existing knowledge about the opioid crisis gathering information with a bottom-up unsolicited approach might be extremely valuable. 

In recent years, Social Media changed the way drug users share information about opiates, give online warnings and avoid potentially toxic drug batches \cite{Guardianreddit},  pointing at ``Reddit: the front page of Internet'' \cite{reddit} as a promising digital source of information. 
Reddit is a social content aggregation website, the 5th most popular website in the US (\texttt{Alexa.com}, Fall 2018),
on which users can
post, comment and vote content.
It is structured in cross-referenced yet independent sub-communities (i.e. \textit{subreddits}), centered around a variety of topics ranging from 
science to drugs and porn
\cite{medvedev2018anatomy}. Reddit is constantly growing, with a total of almost 800 million comments in the year 2016 and almost a billion comments in the year 2017.
Given the ease to register with a ``throwaway'' account, 
Reddit is often used to discuss topics otherwise considered socially unacceptable or unsuitable for the mainstream; users can actively  engage with others, talking uninhibitedly about personal experiences \cite{manikonda2018twitter}, receiving back social support and even life-saving tips from sub-communities of peers \cite{Guardianreddit}.
However, navigating such massive platforms and finding areas of specific interest is usually cumbersome since topics are self-organized bottom-up through users' interactions, leaving the     users to find relevant topics by word of mouth or using a basic search feature. 

The main contributions of this work are summarized as
follows:
\begin{itemize}
\item Design a general purpose information retrieval algorithm able to identify regions of interest when conducting epidemiological surveillance and monitoring on social media

\item Provide an open domain specific vocabulary related to opiates discussions

\item Demonstrate how information disclosed by Reddit users can be used to estimate their geographical location 

\item Identify a novel digital cohort of individuals addressing in a pseudonymous fashion health related topics

\item Provide prevalence maps of interest in opiates consumption

\end{itemize}

\section{Related work}\label{sec:relatedwork}

\textit{Digital epidemiology}~\cite{salathe2012digital}, also referred to as \textit{digital disease detection}~\cite{brownstein2009digital} and \textit{infodemiology}~\cite{eysenbach2009infodemiology},
broadly includes the use of the Internet and digital technologies for collecting and processing health information, both in aggregate form for public health surveillance or from individuals for personal health monitoring. 
Data from a so called \textit{digital cohort} might be collected with active participation of individuals, as in the case of participatory epidemiology \cite{freifeld2010participatory}, but might also be collected passively, e.g. from Social Media, as a byproduct of platforms designed for different purposes. Participatory systems have been implemented through the use of Web platforms~\cite{paolotti2014web} and signals collected from such systems have been shown to be useful for epidemic forecasting~\cite{zhang2017forecasting}.
Data collected from the Web and Social Media have also shown to be useful for monitoring different infectious diseases~\cite{kass2013social,chunara2012social,brownstein2009influenza,brownstein2017combining,sharma2017zika}.

Reddit 
has already proven to be suitable for a variety of research purposes, ranging from the study of 
user engagement 
and interactions between highly related communities~\cite{tan2015all,hessel2016science} to post-election political analyses \cite{barthel_2016}.  Also, it has been useful to study the impact of linguistic differences in news titles  \cite{horne2017impact} and  to explore recent web-related issues such as hate speech \cite{saleem2017web} or cyberbullying \cite{rakib2018using} as well as health related issues like mental illness \cite{de2014mental}, also providing insights about the opioid epidemics~\cite{park2017towards}.
It is  worth to notice that since the platform is collectively generated by the users as in most Social Media, there is not a blueprint of Reddit to perimeter the area under study, and only few attempts tried to overcome this issue extracting meaningful maps or suitable embedding spaces \cite{olson2015navigating,martin2017community2vec}.

We approach the problem of selecting subreddits that are relevant for a specific topic as an information retrieval (IR) problem, where it is possible to retrieve topic-specific documents by expressing a limited set of known keywords. 
Language models~\cite{ponte1998language,schutze2008introduction,croft2013language} tackle this problem using a probabilistic approach with 
the idea that words that are relevant for a given topic would be more likely to appear in a relevant document.
While language modeling is not a common approach for ranking documents collected from Social Media due to the inherent sparsity of the documents --
e.g. for Twitter, more elaborated IR approaches are needed to resolve the sparsity of short texts for tweets~\cite{naveed2011searching} --
subreddits, on the other hand, can be seen as very rich documents from which topic-specific word distributions can be built.

The set of keywords expressed by the user may not include some yet unknown terms that are relevant to the topic but very specific to the language models of the documents, pointing the necessity of query expansion techniques.
Relevance feedback~\cite{rocchio1971relevance} and pseudo-relevance feedback~\cite{buckley1995automatic} are common approaches for query expansion; many of these approaches use human judgment to manually select a subset of the top retrieved documents, and use them to expand the query. More recently, word embeddings~\cite{kuzi2016query} have been used to expand the query with terms that are semantically similar to the query terms.
Techniques that are based on language modeling 
might incorporate term proximity information~\cite{ermakova2016proximity}
to address the automatic query expansion problem, or use an approach based on Information Theory, exploiting the \textit{Kullback-Leibler distance}  for query reweighing \cite{carpineto2001information} and for training local word embeddings \cite{diaz2016query}. 

\section{Dataset and data preparation}\label{sec:dataset}

A dataset~\footnote{\url{https://files.pushshift.io/reddit/}} containing the list of all submissions and comments published in Reddit since 2007 is publicly available online ~\cite{Baumgartnerreddit} and is maintained monthly by adding recent entries.
The dataset is not 100\% complete and it contains some gaps
~\cite{gaffney2018caveat}, 
but these are very small for the years 2016 and 2017 (around 1\% missing data per month).
We use the union of all submissions and comments from years 2016 and 2017, 
for a total of 1,980,497,553 entries. Only subreddits with at least 100 entries 
were selected, resulting in a set of 
1,973,863,886 entries with 74,810 distinct subreddits and 15,747,502 distinct users. 
The text of each entry is parsed and tagged using the spaCy NLP library~\footnote{\url{https://spacy.io/}} v1.9.0. 
For the part-of-speech tagging, a greedy averaged perceptron model was used~\cite{honnibal2015improved}.
Finally, lemmatization is applied to each POS tag; the English lemmatization data is taken from WordNet~\cite{miller1995wordnet} and lookup tables are taken from Lexiconista~\footnote{\url{http://www.lexiconista.com/datasets/lemmatization/}}. After all terms are lemmatized, we select those that appear at least 100 times in the corpus, resulting in a vocabulary of 762,746 lemmas.

\section{identifying relevant documents via document ranking and query expansion}\label{sec:opiates_authors}

As discussed in Section \ref{sec:relatedwork}, the wealth of information contained in Reddit data is not readily available and has to be thoroughly mined. In this section we describe an iterative methodology of semi-automatic retrieval of documents in heterogeneous corpora, in which human intervention is as little as possible.  It is worth to stress that the approach is general and fully unsupervised. However, we added a human-in-the-loop to include domain expert knowledge in the process and reach better results. On the other hand, a domain expert alone without the aid of the algorithmic pipeline for document ranking and query expansion  would have been hopeless in navigating the Reddit world by hand.
The steps are summarized in Algorithm \ref{algo}.
We start with a small set of keywords $Q$ provided by the user, or \textit{query} in the following. At each iteration, we select documents which are both relevant to the query and informative,
and enrich the \textit{query} terms set until we arrive to a stable list of documents and query terms.
While this methodology works well on Reddit in the domain of topics related to the opioid epidemics (see Section \ref{sec:opiates subreddits}), it is also sufficiently general to be used for other information retrieval tasks and might be valuable for different epidemiological research questions.

\begin{algorithm}
\SetKwInOut{Parameters}{Parameters}
  \KwIn{Corpus $C$, query $Q$}
  \Parameters{$n$, $m$, $\alpha$} 
 Initialize the vocabulary $\textbf{V}$\;
 \ForEach{word $w \in \textbf{V}$}{
   calculate $p_C(w)$\;
   \ForEach{document $d \in C$}{
   	calculate $p_d(w)$\;
   }
 }
 $Q_{new} \leftarrow Q$\;
 $K_{new} \leftarrow \varnothing$\;
 \Repeat{$Q = Q_{new}$ and $K = K_{new}$}{
    $Q \leftarrow Q_{new}$\;
 	$K \leftarrow K_{new}$\;
    $R_d$ $\leftarrow$ Rank documents using $\mathrm{score}(d \mid Q,C)$ (Eq. \ref{eq:kld})\;
    $K_{new}$ $\leftarrow$ top $n$ documents in $R_d$\;
    $R_w$ $\leftarrow$ Rank terms using $\mathrm{score}(w \mid K_{new})$ (Eq. \ref{eq:loglikelihood})\;
    $Q_{candidate}$ $\leftarrow$ top $m$ terms from $R_w$\;
    $Q_{new}$ $\leftarrow$ manual selection of terms in $Q \cup Q_{candidate}$\;
 }
 \KwOut{$R_d$, $R_w$}
 \caption{IR steps for document ranking}
 \label{algo}
\end{algorithm}

First, we create a general vocabulary $\textbf{V}$ by collecting terms from the entire corpus in a bag-of-words fashion. 
We compute the probability of occurrence of a term $w$  in the entire corpus $C$ as the ratio  $p_C(w) = f_C(w)/\sum\limits_{w} f_C(w) $ between its raw count  $f_C(w)$ in the corpus and the total number of words in the corpus. Let us also define the regularized marginal probability of occurrence of term $w$ in a document $d$ as
\begin{equation}\label{eq:probability}
p_d(w) = \frac{f_d(w)}{\sum_w f_d(w)  + \alpha} + p_C(w).
\end{equation}
where $f_d(w)$ is the count of $w$ in $d$. In case of corpora with very heterogeneous document sizes,
the regularization term $\alpha$ is added to control ``the size'' of the language model of the documents (small documents will result in small marginal probabilities). In our experiments, we use $\alpha = 10^4$, so documents with total number of words lower than $10^4$ have ``flattened'' probabilities in their language model.
Adding 
$p_C(w)$ to 
the marginal probability of $w$ 
reduces the impact of  
words that are rare or not present in a document,
and only words that are more likely to appear in the document will impact the document ranking.
The use of these two regularization terms ($\alpha$ and $p_C(w)$) are effective in low-count scenarios and have a small impact for words with high probability in the document or in case of documents with large language model.

With the intuition that a document will result to be relevant in the context of the query if it contains query terms more likely to appear in the document than in the general corpus, we evaluate
\begin{equation}
\mathrm{score}(d \mid Q,C) = KLD_Q(p_d,p_C)=\sum_{w \in Q} p_d(w) log\frac{p_d(w)}{p_C(w)}
\label{eq:kld}
\end{equation}
which is the total contribution of the query terms in the \textit{Kullback-Leibler divergence} between the document and the whole corpus $C$.\\
We consider the top $n$ documents ranked by relevance as measured by Eq. (\ref{eq:kld}) as the \textit{set of relevant documents} $K$.
Once $n$ is chosen (and thus $K$ is obtained), in order to enrich the query terms $Q$ we assign a score to each term based on the logarithm of the  likelihood ratio 

\begin{equation}
\mathrm{score}(w \mid K) = \sum_{k \in K}log\frac{p_k(w)+p_C(w)}{p_C(w)}.
\label{eq:loglikelihood}
\end{equation}

Whenever the the maximum likelihood estimate $p_k(w)$ of a term in a relevant document is higher than its estimate in the general context $p_C(w)$ -- i.e. the term is more informative in the
document than in the general context -- the contribution to the score will be high and positive. Conversely, whenever a term is less likely to appear in the document than in general context,
its contribution to the score will be smaller,
highlighting that the term is common or not relevant to the context.
Considering the top $m$ terms ranked by Eq. (\ref{eq:loglikelihood}), a subset of previously unknown relevant terms is added to $Q$ with the supervision of an expert.
With the newly enriched set of query terms, the entire pipeline can be iteratively evaluated until convergence, reached if no new documents are added to the top $n$ documents and no new relevant terms among the top $m$ can be added to $Q$.
These steps are summarized in Algorithm \ref{algo}, and the document ranking resulting from the last iteration of the algorithm is used to select the most relevant documents.

\begin{figure}
\includegraphics[width=0.93\columnwidth,trim={0cm 0cm -0.2cm 0cm},clip]{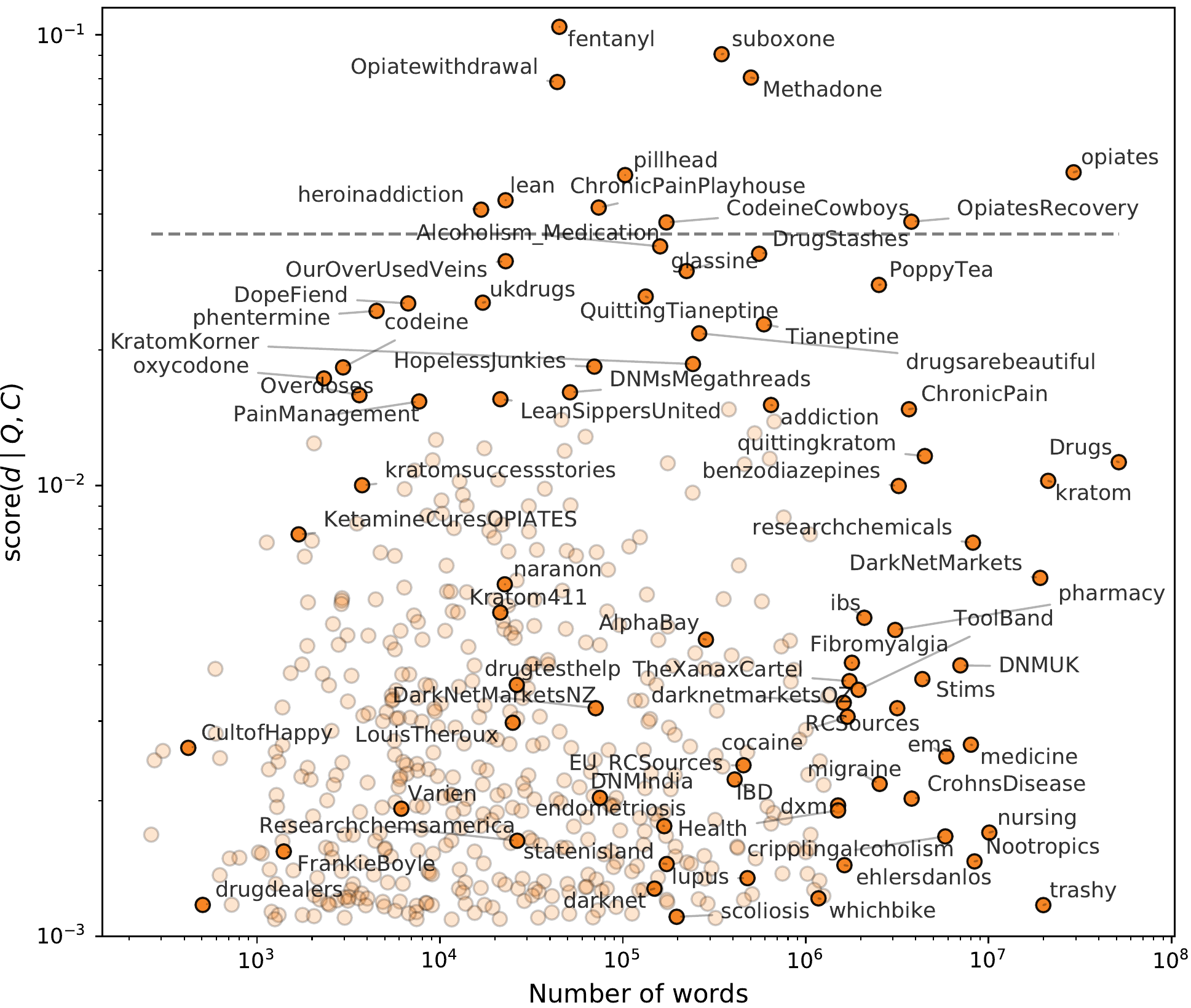}
 \includegraphics[width=0.93\columnwidth,trim={-0.2cm -0.1cm 0cm 0cm}, clip]{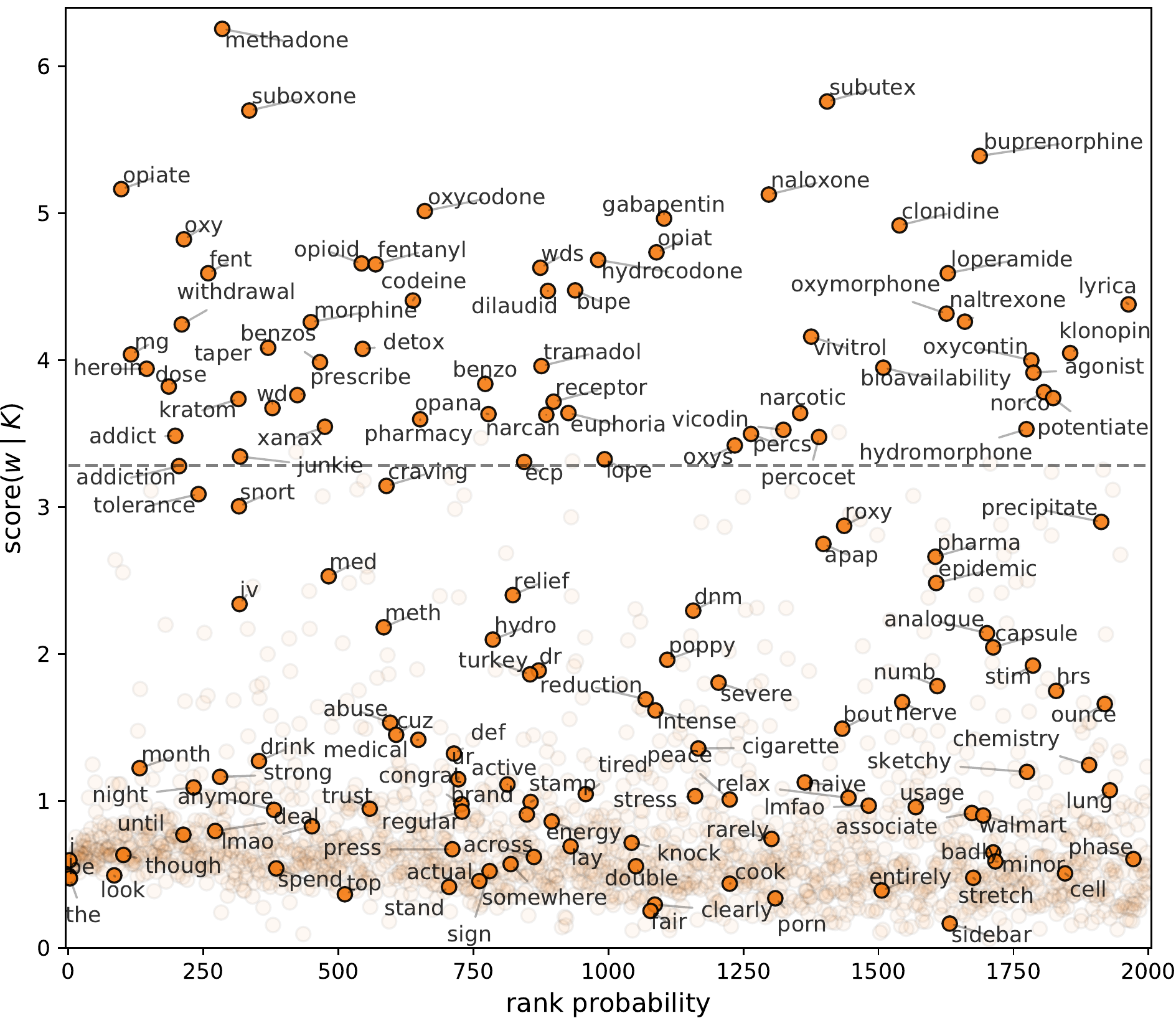}
\caption{\textit{Opiates subreddits} (top): individual subreddits are shown in a coordinate space of the number of words in the subreddit and the final ranking score. The subreddits above the dashed line are those selected as $K$ opiates related subreddits.
\textit{Opiates vocabulary} (bottom): Top 2,000 terms sorted by rank probability of term in the set $K$. Terms above the dashed line were selected as query term candidates in the last step of the subreddit retrieval algorithm.}
 \label{fig:scatter_clean_vocabulary}
\end{figure}

\subsection{Opioid related subreddits}\label{sec:opiates subreddits}

Our assumption is that authors who post
content
in a subreddit related to a particular topic are \textit{interested} in that topic. Therefore we consider all authors participating in threads on subreddits related to opioid consumption as those \textit{interested} in the topic, and we discover such subreddits 
by applying the algorithm described in Section \ref{sec:opiates subreddits}.
After data preparation steps described in Section \ref{sec:dataset},
starting from a list of opiates related keywords 
$q =$ \texttt{[fentanyl, oxycodone, suboxone, morphine]} 
we extract the top $n=10$ subreddits:
\texttt{[suboxone, fentanyl, Opiatewithdrawal, TarkovTrading, heroinaddiction, ChronicPainPlayhouse, OpiatesRecovery, opiates, Methadone, PoppyTea]}.
We then proceed with the iterative procedure of query enrichment and document ranking, considering as
relevant
terms only 
opioid drug names,
i.e. chemical names (e.g. \textit{oxycodone}), brand names (e.g. \textit{Percocet}) and street slang (e.g. \textit{percs}), disregarding drugs that might be abused together with opiates (like benzodiazepines) but are not in the opiates domain.
The final set of \textit{opioid related subreddits} used in this paper is $K = $ \texttt{[fentanyl, suboxone, Opiatewithdrawal, Methadone, opiates, pillhead, lean, ChronicPainPlayhouse, heroin-\linebreak addiction, OpiatesRecovery]}. All the 37,009 users who posted on such subreddits in 2016 or 2017 are then considered as interested in opiates.
Figure~\ref{fig:scatter_clean_vocabulary}~(top) shows the final ranking score for the subreddits, plotted against their size in terms of number of words. 

\subsection{Opioid specific vocabulary}

 When applied to Reddit, the query expansion approach is particularly useful in revealing how a topic is discussed on subreddits. 
Given the large user base of Reddit and the tendency of the users in employing slang and street names alongside proper drug names, 
this method is very helpful in acknowledging 
alternative names of drugs,
like \textit{sub} for \textit{suboxone} and \textit{bth} for \textit{black tar heroin}.

As a byproduct of the proposed methodology applied to the opiates domain, we extracted a topic-specific vocabulary by weighting each term of the vocabulary with 
Eq. \ref{eq:loglikelihood}.
Very specific opioid-related terms (i.e. with high probability in opiates subreddits and low probability in the whole corpus) have large positive values of $score(w \mid K)$, as shown in Figure~ \ref{fig:scatter_clean_vocabulary}~(bottom),  while stop-words and common terms 
have small score values.
A total of 2,616 terms out of the original 762,746 (0.3\%) have score higher than 1.
The full list of vocabulary terms ranked by score is  available for research purposes~\footnote{\url{https://github.com/ISIFoundation/WWW19OpiatesAbuseSocialMedia}}.

\section{Geolocating users on Reddit}\label{sec:geolocating}

Reddit does not provide any explicit information about users' location, therefore we apply three methodologies to assign a location to users. Finally, we merge the mappings in a single user-location matching.

\textbf{1. Self reporting via regular expression}: Reddit users often declare geographical information about themselves in submission or comment texts.
We selected all texts containing the expression \texttt{`I live in'} 
(3,337,850 instances in 2016 and 2017)
and extracted candidate expressions from the text that follows, to identify which ones represented US states and cities.
We started with a set of US cities from the GeoNames database~\footnote{\url{http://www.geonames.org/}} with population higher than 20k, and selected only the candidate expressions that included both the city name and the state (e.g. `Newark, California' or `Newark, CA') to avoid confusion with cities with same name (e.g. `Newark, New Jersey'). Once the US state for these expressions were assigned and removed from the candidate expressions, we proceeded to all US cities with population higher than 200k, selecting expressions with the name of the city and their variants (e.g. `New York', `Big Apple'). After assigning the corresponding US state for these expressions and removing them from the candidates, we proceeded to select the expressions with a state name on it (e.g. `Alabama', `California'). Among the initial set of candidate expressions, 
886,919 (27\%) had a state associated to it.
By removing inconsistent self reporting (13,374 users who reported more than one US state) we geolocated 378,898 distinct users.
 
\textbf{2. Self reporting via \textit{user flairs}}: In Reddit, \textit{user flairs} are attributes (usually selected by the users) that are attached to their submissions or comments in a specific subreddit. In some subreddits flairs might be limited to a set of geographical locations (countries, states, cities and  city neighborhoods), meaning that users should identify themselves with one of these locations.
A user selecting a location flair is therefore considered equivalent to a user self reporting its location.
We mapped the users participating in subreddits  with location flairs referring to US states to their \textit{flaired} positions.
Using this approach, we mapped 206,125 users 
to the 51 US states (including District of Columbia) by selecting the most common among the position flairs expressed by a user.

\textbf{3. Posting on subreddits specific to locations}: Reddit includes subreddits 
discussing topics specific to geographical locations
(e.g. \texttt{r/Alabama} or \texttt{r/Seattle}).
The subreddit \texttt{r/LocationReddits} \linebreak keeps 
a curated
list of these local subreddits.
We collected from the page corresponding to North America\footnote{\url{https://www.reddit.com/r/LocationReddits/wiki/faq/northamerica}} the mappings of 2,844 subreddits to 51 US states.
By assuming that a user who posts comments in one of these subreddits is likely to live that location, we estimated the position of 1,198,096 authors.

After retrieving US states positions using the three methods above, we found that about 12\% of mapped users expressed multiple locations. In order to uniquely map authors and states, for location flairs and LocationReddits sources we assigned each author a unique location, it  being the most frequent among the ones expressed by the author.
We discarded authors whose most frequent location was not unique.
This resulted in 194,008 authors retrieved via location flairs (5.9\% loss) and 1,077,516 via LocationReddits (1.4\% loss).

\begin{figure}
\includegraphics[ width=0.9\columnwidth,trim={0.8cm 1cm 1.5cm 2cm},clip]{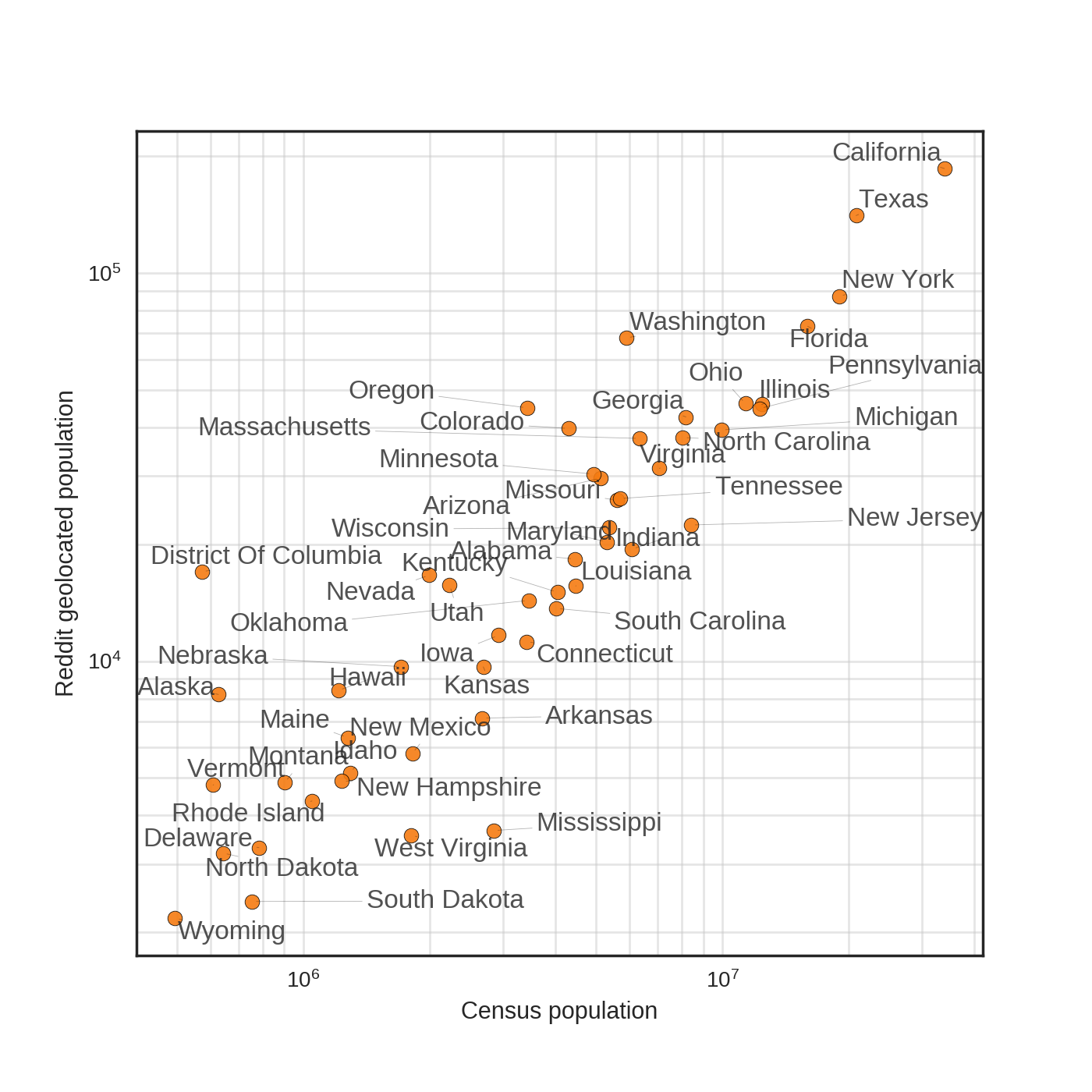}

\caption{ \textit{Reddit geolocated population}: scatter plot of the number of geolocated Reddit users and census population.
}
\label{fig:relative_mismatch}
\end{figure}

We evaluated Pearson's $r$ correlation between the log of the 2000 US
Census population and the log of the population assigned to the same US states using the three methodologies. Results are in good agreement for all sources, with $r$= 0.85, 0.91 and 0.86 for respectively User flairs, Regular expression and LocationReddits,
and all $p$-values below 1e-12.

Finally, we merged the information from all three sources in a unique location for each author.
We considered the regular expression technique to be the most reliable due to its unambiguous self reporting nature, resulting in the highest correlation with census data. 
We proceeded in the merging process by first assigning the authors their regular expression location, if present. If missing, we assigned them their position from the joint information of location flairs and LocationReddits by summing the occurrences of locations expressed in the two sources and verifying the uniqueness of the most frequent location.
 Although some approaches have been proposed to geolocate users using language models \cite{han2014text}, we rely on a conservative approach with the aim of reducing misclassification, considering only explicit geographical information directly provided by the authors.
 The full set of users geolocated using the above methodology consists of 1,408,388 users, 
with state representativeness in the order of 5.5 Reddit users per thousand U.S. residents (median value among all U.S. states).
Although we acknowledge a potential bias due to heterogeneities in Reddit population coverage and users demographics, the number of Reddit users has good linear correlation of $r$ = 0.89 and $p$-value below 1e-12 with census population (Figure~\ref{fig:relative_mismatch}).

\section{Prevalence of interest in opiates}\label{sec:prevalence}

\begin{figure*}[!ht]
\includegraphics[width =0.74\paperwidth ,trim={2.2cm 0.9cm 0cm 0.1cm},clip]
{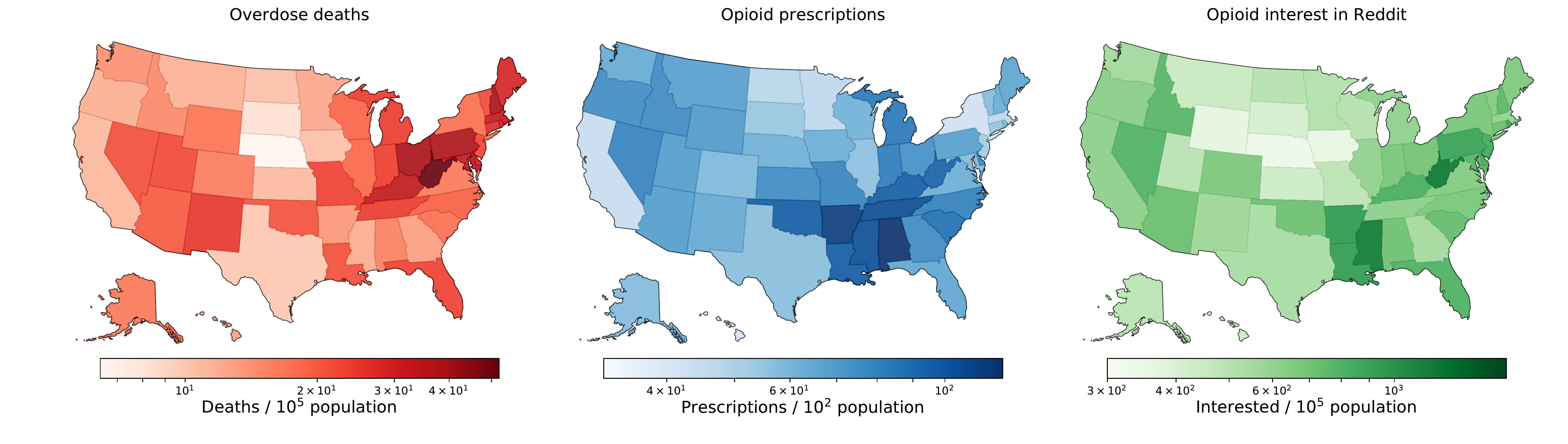}
\centering
\caption{\textit{US states distribution maps:} choropleth maps representing the overdose deaths rate for 2016 (red), the opioid prescription rate for 2016 (blue), the opioid interest rate in Reddit for 2016 and 2017 (green).}
 \label{fig:maps_prevalence_regression}
\end{figure*}

Conversations in opiates related subreddits branch off in many topics, mostly regarding opiates usage, dosages, interactions with other substances, safe practices and withdrawal, usually with a personal perspective. Users share their health and addiction status and provide support among each other. In general, they share a common and firsthand interest in opiates experiences.
Thus, the number of authors participating in the conversations in opiates related subreddits (as identified in Section \ref{sec:opiates_authors}) is not to be considered as a crude number of opiates users and addicts but it is rather to be considered as a proxy of users personally interested in opiates in the broadest sense. 

Using the geographical information of Reddit population estimated in Section \ref{sec:geolocating} and having identified the opiates authors in Section \ref{sec:opiates subreddits}, we were able to evaluate the \textit{opiates interest prevalence} at the US state level as the fraction of geolocated users engaged in opiates subreddits and the total estimated population.

We mapped to US states the 9,026 geolocated users who posted in opioid related subreddits, equivalent to 24\% of the opiates authors, for a mean interest prevalence of 636 per 100,000 Reddit users (CDC data for 2016 reported an age-adjusted rate of overdose deaths of 19.8 per 100,000 , and opioid prescription rate of 66.5 prescriptions dispensed per 100 persons).
\setlength\tabcolsep{4pt}
\begin{table}
\begin{tabular}{llrrr}
\toprule
{} &{} &   Opiates & Reddit & Interest\\
{} & {} &  authors & authors& prevalence\\
Region & Division           &                     &                    &                     \\
\midrule
Northeast & Middle Atlantic    &                1,186&           154,418&     768.05\\
Northeast & New England        &                 455&            69,132&     658.16\\
Midwest & East North Central &                1,082&           172,902&     625.79\\
Midwest & West North Central &                 424&            92,931&        456.25\\
South  & East South Central &                 457&         63,269&        722.31\\
South & South Atlantic     &                1,656&        242,470&        682.97\\
South  &  West South Central &                1,079&           177,856&        606.67\\
West  &  Mountain           &                 793&           119,742&        662.26\\
West &  Pacific            &                1,894&           315,668&        599.10\\
\bottomrule\\
\end{tabular}
\caption{\textit{Interest prevalence by US regional division}:  number of opiates authors, total number of authors, and interest prevalence per 100,000 individuals measured on Reddit.
}
\label{tab:division_prevalence}
\end{table}
The areas of greater interest according to our estimates is the South Region (Table \ref{tab:division_prevalence}), with high prevalence for Mississippi, Arkansas, Louisiana and the highest measured value of 1,180 interested users per 100,000 population in West Virginia (Figure \ref{fig:maps_prevalence_regression}, green map).  Middle Atlantic and New England states like Pennsylvania, New Jersey and Rhode Island are also largely involved, showing high interest rates ranging between 850 to 900 individuals per 100,000.
In line with official statistics about drugs overdose deaths, West North Central states are those with the lowest interest rate measured on Reddit, ranging from 341 per 100,000 in  Nebraska to 510 per 100,000 in Minnesota.

We confronted the estimated interest prevalence with official statistics from the Centers for Disease Control and Prevention~\footnote{\url{https://www.cdc.gov/drugoverdose/data/index.html}} grasping different angles of the opioid crisis. In particular we focused on opioid drug overdose deaths rates and retail opioid prescribing rates, both regarding 2016 (the most recent available dataset at the time of writing), shown in Figure \ref{fig:maps_prevalence_regression} in red and blue respectively.
These two phenomena seem fairly uncorrelated, with a Pearson's correlation of 0.068 (p-value of 0.637). It is worth to stress that those ``gold-standard'' data are the only ones provided by CDC that allows for comparisons between different states: although the counting of drugs overdose deaths includes every drug and is not broken down by drug type, state-level estimates of opiate-related overdose deaths are affected by heterogeneities in the surveillance system. On the other hand, official statistics about prescribing rates that includes both appropriate prescriptions and drug abuse are affected by different prescription policies in place at different states.

The interest prevalence shows fairly high positive linear correlations with  CDC rates, respectively r = 0.45 (p-value = 8.4e-04) with the opioid overdose deaths rate and r = 0.506 (p-value = 1.6e-04) with the retail opioid prescribing rate.
These correlations suggest that the signal of interest in opiates measured on Reddit partially explains the observed phenomena around opioid epidemics measured by the standard surveillance system.
Moreover, we trained a linear regression model to fit the estimated prevalence using drug overdose death rates and prescriptions rates as features and predicted new values of interest prevalence, 
resulting in an higher correlation of r = 0.655 (p-value = 1.7e-07) with the estimated interest prevalence.
This result confirms that we are sensing a broad signal, tied to drug overdose deaths and opioid prescriptions but probably accounting for more complex aspects of the phenomenon.

\begin{figure}[ht]
\centering
\includegraphics[width=0.9	\columnwidth]{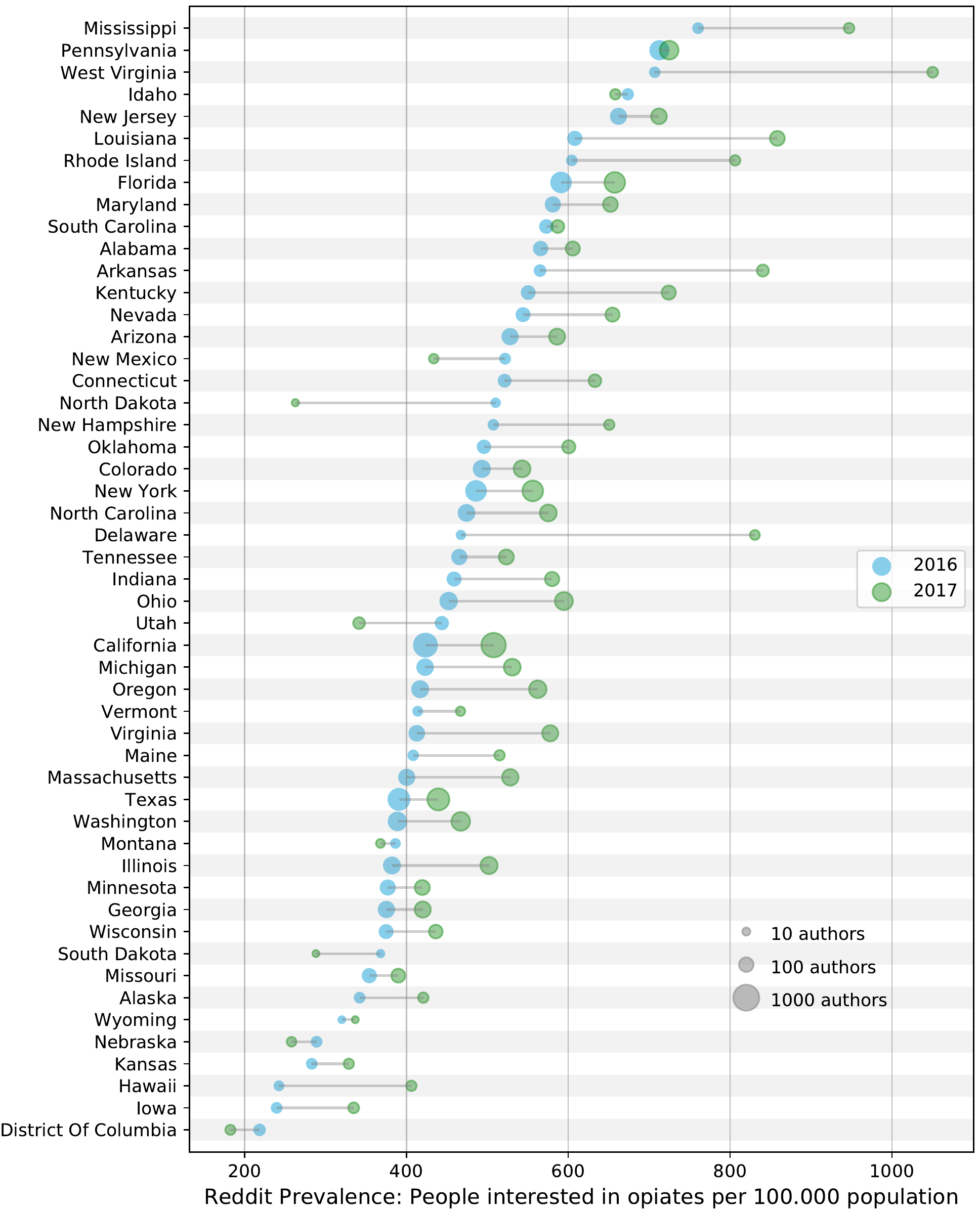}
\caption{\textit{Opioids interest prevalence:} number of Reddit authors per 100,000 Reddit population. 
Prevalence values for years 2016 and 2017 are reported on the x-axis. The size of the bubbles is proportional to the number of Reddit authors.}
 \label{fig:prevalence_2016_2017}
\end{figure}

Leveraging the geolocated cohort, we also evaluated the temporal variation of interest prevalence between 2016 and 2017 broken down by state.
According to 
Figure~\ref{fig:prevalence_2016_2017}, the interest prevalence decreased only in 8 states while in general we observe that in areas with a good coverage of opiates-related users (namely, California, Texas, New York, Florida), the interest prevalence increased by 10\% to 20\%. It is worth to stress that, at the time of writing, no official data about 2017 drug overdose deaths and associated trends are available, highlighting the tremendous potential of a digital epidemiology approach to gather timely insights about hard-to-reach information of health-related topics at the population level.

\section{Conclusion}
This study provides an analysis of Reddit content related to personal opiates abuse in the period of 2016 and 2017. Starting from almost 2 billion
posts over 74k distinct subreddits, we applied a general information retrieval algorithm to identify specific subreddits of interest thus selecting 37,009 users that show an explicit interest in the topic. 1.5 million pseudonymous Reddit users were geolocated at US state level by 
looking at the content they generated in the platform.
The number of mapped users for each state are in good agreement with census data, with some differences in terms of coverage.
Such cohort might represent a biased yet valuable digital observatory on several social, political and health-related topics. 
The prevalence of opiates interest extracted with the presented approach shows a complementary perspective to official surveillance, and its geographical heterogeneity partially encodes signals from opioid prescribing rates and drug overdose deaths.

\begin{acks}

  DB acknowledges support from the Lagrange Project and CRT Foundation (\url{http://isi.it/en/lagrange-project/project}).

\end{acks}

\pagebreak
\bibliographystyle{ACM-Reference-Format}
\bibliography{bibliography}

\end{document}